\documentclass[twocolumn,prd,superscriptaddress,showpacs,floatfix,%
preprintnumbers,nofootinbib]{revtex4}

\usepackage{epsfig}
\usepackage{bm}

\usepackage[dvips]{color}
\definecolor{Black}{named}{Black}
\definecolor{Blue}{named}{Blue}
\definecolor{Red}{named}{Red}

\newcommand{\gag}{g_{a\gamma}}
\long\def\dump#1{}

\newcommand\I{{\rm i}}

\begin{document}

\title{Effects of Axion-Photon Mixing on Gamma-Ray Spectra from
Magnetized Astrophysical Sources}

\author{Kathrin A.~Hochmuth}
\affiliation{Max-Planck-Institut f\"ur Physik
(Werner-Heisenberg-Institut), F\"ohringer Ring 6, 80805 M\"unchen,
Germany}

\author{G{\"u}nter Sigl}
\affiliation{APC~\footnote{UMR 7164 (CNRS, Universit\'e Paris 7,
CEA, Observatoire de Paris)} (AstroParticules et Cosmologie),
10, rue Alice Domon et L\'eonie Duquet, 75205 Paris Cedex 13, France\\
and Institut d'Astrophysique de Paris, UMR 7095 CNRS - Universite Pierre \&
Marie Curie, 98 bis boulevard Arago, F-75014 Paris, France}

\begin{abstract}
Astrophysical $\gamma-$ray sources come in a variety of sizes and magnetizations.
We deduce general conditions under which $\gamma-$ray spectra from such
sources would be significantly affected by axion-photon mixing. We show that,
depending on strength and coherence of the magnetic field,
axion couplings down to $\sim(10^{13}\,{\rm GeV})^{-1}$ can give rise
to significant axion-photon conversions in the environment
of accreting massive black holes. Resonances can occur between the axion
mass term and the plasma frequency term as well as between the plasma frequency
term and the vacuum Cotton-Mouton shift. Both resonances and non-resonant
transitions could induce detectable features or
even strong suppressions in finite energy intervals of $\gamma-$ray spectra
from active galactic nuclei. Such effects can occur at keV to TeV energies for 
couplings that are currently allowed by all experimental constraints.
\end{abstract}



\maketitle

\section{Introduction}
Axions and axion-like particles couple to two photons and can thus
convert into on-shell real photons and vice-versa in the presence of
magnetic fields. Astrophysical sources produce high energy
$\gamma-$rays as secondary particles of primary charged cosmic rays.
Since these primary cosmic rays have to be accelerated in electromagnetic
fields, the spectrum of any astrophysical $\gamma-$ray source can
potentially be modified by axion-photon mixing. In the present paper
we discuss in detail in which source environments and for
which axion masses and coupling constants one can expect significant
modifications of the observable $\gamma-$ray spectrum. As specific
examples we consider the jets and central engines of active galactic
nuclei (AGN) both of which are driven by magnetized accretion disks.

We consider the Lagrangian for the coupling between a pseudo-scalar
field $a$, called axion in the following, and the electromagnetic
field strength $F_{\mu\nu}$
\begin{equation}\label{eq:lagrangian}
  {\cal L}_{\gamma a}=-\frac{1}{4}g_{\gamma a}
  F_{\mu\nu}\tilde{F}^{\mu\nu}\,a=g_{\gamma a}{\bf E}\cdot{\bf B}\,a\,,
\end{equation}
where $\tilde{F}_{\mu\nu}\equiv\frac{1}{2}\varepsilon_{\mu\nu\rho\sigma}
F^{\rho\sigma}$ is the electromagnetic dual, ${\bf E}$ and ${\bf B}$ are
the electric and magnetic field strengths, respectively, and $g_{\gamma a}$
is the photon-axion coupling. Eq.~(\ref{eq:lagrangian}) implies that
a photon can convert into an axion in a magnetic field ${\bf B}$.
In the present work we will consider the effect of such conversions
on $\gamma-$ray fluxes emitted from magnetized sources for general
pseudo-scalar masses $m_a$ and photon-axion couplings $g_{\gamma a}$.

Axions with couplings $g_{\gamma a}\lesssim10^{-10}\,{\rm GeV}^{-1}$
are consistent with all existing constraints for almost all axion
masses~\cite{Battesti:2007um} and we will use $g_{\gamma a}=10^{-11}\,{\rm GeV}^{-1}$ as our benchmark value. As possible $\gamma$-ray sources we consider already detected sources, like Makarian 501 and Makarian 421. Sources of this type can have a model-dependent magnetic field strength between 0.01 G and several G. Adopting these values we find that a significant amount of $\gamma$-rays can be converted into axions, thereby leaving characteristic signatures like steps or gaps in the spectrum. This conversion can take place due to normal oscillations and due to resonance effects. We find that for typical parameters the effects can be sizable and are well within the detection region of the next generation of experiments.

Our paper is organized as follows. In Sec.~\ref{sec:osc} we discuss in general how the $\gamma$-ray spectra can be modifed due to resonant and non-resonant oscillation effects. We discuss the prospects of detecting these effects in Sec.~\ref{sec:sources} where we discuss specifically the case
of central engines and jets of AGN. In Sec.~\ref{sec:conclusions} we present
our conclusions. We use natural units, $\hbar=c=k=1$, throughout the paper.

\section{Conditions for Modification of gamma-ray Spectra}
\label{sec:osc}
The axion-photon conversion probability in a transverse magnetic field $B_{\rm t}$ can be derived by using the linearized equation of motion following from Eq.~(\ref{eq:lagrangian}) for relativistic axions. This equation can be written as~\cite{Raffelt:1987im}
\begin{equation}  \label{linsys}
\left(E-{\I}\partial_z-{\cal M}\right) \left(
\begin{array}{ccc}
A_{\perp}\\
A_{\parallel}\\
a
\end{array}
\right)=0\,,
\end{equation}
where $z$ is the direction of propagation, $E$ is the photon energy and $a$ the axion field. $A_\perp$ and $A_\parallel$ are orthogonal components of the photon field, where $i={\perp}$ or ${\parallel}$ refer to the $B_{\rm t}$ direction. The mixing matrix ${\cal M}$ is 
\begin{equation}  \label{mixmat}
{\cal M}\equiv\left(
\begin{array}{ccc}
\Delta_{\perp} &\Delta_{\rm R}     & 0\\
\Delta_R       &\Delta_{\parallel} & \Delta_{B}\\
0              &\Delta_{B}   & \Delta_a\\
  \end{array}
  \right)\,,
\end{equation}
where
\begin{equation}\label{eq12}
\begin{array}{cccccc}
\Delta_{\perp}&=&\Delta_{\rm pl}+\Delta_{\rm CM}^{\perp}\,,& \quad
\Delta_{B}&=& \frac{1}{2}\gag B_{\rm t}\,,\\
\Delta_{\parallel}&=&\Delta_{\rm pl}+\Delta_{\rm CM}^{\parallel}\,,
&\quad \Delta_{\rm pl}&=& \omega_{\rm pl}^2/(2E)\,.
\end{array}
\end{equation}
Here we have defined $\omega_{\rm pl}^2 = 4\pi\alpha\,n_e/m_e$ as the plasma
frequency for an electron density $n_e$, where $m_e$ is the electron mass and
$\alpha\equiv e^2/(4\pi)$ the fine-structure constant with $e$ the electron charge.
$\Delta_{\rm R}$ is the Faraday rotation term, which is dependent on the energy of the longitudinal component of the magnetic field and couples to the modes $A_{\parallel}$ and $A_{\perp}$. However, we want to consider only non-polarized sources, which renders this term negligible. The vacuum Cotton-Mouton effect is represented by the $\Delta_{\rm CM}$ terms, which describe the birefringence of fluids in presence of a longitudinal magnetic field, with $|\Delta_{\rm
CM}^{\parallel}-\Delta_{\rm CM}^{\perp}|\propto B_{\rm t}^2$. Note that the
photon dispersion relation for polarization $i$ is $E_i(k)=k+\Delta_i$, corresponding
to the refractive indices $n_i-1\simeq\Delta_i/E$. For axions one has, of course,
$E(k)\simeq k+m_a^2/(2k)$ and thus $\Delta_a\simeq m_a^2/(2E)$.

Neglecting different photon polarization states and denoting the
resulting photon state by $A$, we are left with a simple
two-component mixing problem,
\begin{equation}
 \left[E-{\I}\partial_z -
 \left(
  \begin{array}{cc}
    \Delta_{\rm pl}(z)+\Delta_{\rm CM}(z)&\Delta_B(z)\\
    \Delta_B(z)&\Delta_a
  \end{array}
 \right)\right]
 \left(
 \begin{array}{cc}
 A\\a
 \end{array}
 \right)=0\,,
\end{equation}
where we have indicated the terms that are location dependent.
After diagonalization of the mixing matrix we obtain the solution
\begin{equation}\label{tan}
 \theta(z) = \frac{1}{2}
 \arctan\left(\frac{2\Delta_B(z)}{\Delta_{\rm pl}(z)+\Delta_{\rm CM}(z)-\Delta_a}\right)\,.
\end{equation}

Within a domain of linear size $s$ and roughly constant plasma density and
magnetic field, the probability of a transition from a photon to an axion
is~\cite{Raffelt:1987im,Mirizzi:2006zy,Mirizzi:2007hr}
\begin{equation}\label{eq:p_0}
  P_{\gamma\to a}\simeq(\Delta_B s)^2
  \frac{\sin^2\Delta_{\rm osc} s/2}{(\Delta_{\rm osc} s/2)^2}
  \equiv P_0\,,
\end{equation}
where the oscillation wave number is
\begin{equation}\label{eq:Delta}
  \Delta_{\rm osc}^2\simeq(\Delta_{\rm CM}+\Delta_{\rm pl}-\Delta_a)^2+4\Delta^2_B\,.
\end{equation}
In Eq.~(\ref{eq:Delta}), the plasma contribution $\Delta_{\rm pl}$, the 
vacuum Cotton-Mouton term $\Delta_{\rm CM}$, the axion mass term $\Delta_a$, and
the off-diagonal mixing term $\Delta_B$ are given by
\begin{eqnarray}\label{eq:frequencies}
	\Delta_a&=&\frac{m_a^2}{2E}\simeq2.5\times10^{-20}\,m_{\mu{\rm eV}}^2
	\left(\frac{\rm TeV}{E}\right)\,{\rm cm}^{-1}\,,\nonumber\\
  \Delta_{\rm pl}&=&\frac{\omega_{\rm pl}^2}{2E}
  \simeq3.5\times10^{-26}\left(\frac{n_e}{10^3\,{\rm cm}^{-3}}\right)
  \left(\frac{\rm TeV}{E}\right)\,{\rm cm}^{-1}\,,\nonumber\\
  \Delta_{\rm CM}&\simeq&-\frac{\alpha}{45\pi}
  \left(\frac{B_{\rm t}}{B_{\rm cr}}\right)^2\,E\\
  &\simeq&-1.3\times10^{-21}\,B_{\rm mG}^2
  \left(\frac{E}{\rm TeV}\right)\,{\rm cm}^{-1}\,,\nonumber\\
	\Delta_B&=&\frac{g_{\gamma a}B_{\rm t}}{2}\simeq
	1.7\times10^{-21}\,g_{11}\,B_{\rm mG}\,{\rm cm}^{-1}\nonumber\,,
\end{eqnarray}
where $E$ is the photon energy and
$B_{\rm cr}\equiv m_e^2/e\simeq4.41\times10^{13}\,$G
is the critical magnetic field strength. Since we
are only interested in orders of magnitude, we neglect different photon
polarization states whose vacuum Cotton-Mouton term have slightly
different pre-factors, and we have averaged over directions,
$\langle B_{\rm t}^2\rangle\simeq B^2/3$. Further,
in Eq.~(\ref{eq:frequencies}) we have used the abbreviations
$B_{\rm mG}\equiv(B_{\rm t}/{\rm mG})$,
$g_{11}\equiv g_{\gamma a}\times10^{11}\,$GeV, and
$m_{\mu{\rm eV}}\equiv(m_a/\mu{\rm eV})$.

Only terms to lowest non-trivial order in $B$ are taken into account
in Eq.~(\ref{eq:frequencies}). The term to next-higher order in $B$
in $\Delta_{\rm CM}$ is suppressed by a factor
$\simeq\chi^2\equiv\left[(E/m_e)(B/B_{\rm cr})\right]^2$~\cite{Adler:1971wn}.
This is negligible for energies $E\lesssim2\times10^{16}\,({\rm kG}/B)\,$eV.
Furthermore, the magnetic field influences the phase space distribution
of the electrons which modifies the plasma term to
$\simeq(E/2)\omega_{\rm pl}^2/(E^2-\omega_{\rm c}^2)$~\cite{Adler:1971wn}, where
$\omega_{\rm c}=eB/m_e=(B/B_{\rm cr})m_e$ is the electron cyclotron frequency
in a non-relativistic plasma. The latter is thus negligible for energies
$E\gtrsim\omega_{\rm c}\simeq10^{-5}\,(B/{\rm kG})\,$eV. Furthermore,
inelastic processes such as pair production and photon splitting in
the magnetic field are suppressed as $\exp(-1/\chi)$ and thus
negligible for $\chi\ll1$~\cite{Erber:1966vv}. Since we are
mostly concerned with magnetic fields $B\lesssim10^4\,$G and photon energies
$E\lesssim\,$PeV, both inelastic processes and higher order corrections
to Eq.~(\ref{eq:frequencies}) can be neglected.

For propagation over $N$ coherence domains the total conversion probability of
photons into axions can be derived as~\cite{Mirizzi:2006zy}
\begin{equation}\label{eq:p}
  P_{\gamma\to a}\simeq\frac{1}{3}\left[1-\exp\left(-3NP_0/2\right)\right]
  \,,
\end{equation} which saturates to $P_{\gamma\to a}\simeq\frac{1}{3}$ for $NP_0\gg 1$.

\subsection{Resonances}
In the following, let us denote the coherence length of the plasma and magnetic field by
$\lambda$ with $\lambda\equiv\lambda/$pc. The coherence
length is of course always smaller than the size of the system.

MSW type resonances can occur in Eq.~(\ref{eq:frequencies}) 
when $\Delta_{\rm pl}$ becomes comparable to $\Delta_a$
or the modulus of $\Delta_{\rm CM}$ and the other contribution is negligible. In
order for the resonance to lead to efficient conversion, it has to
be adiabatic, requiring
$2\pi\left|\Delta_{\rm pl}^\prime+\Delta_{\rm CM}^\prime\right|\lesssim\Delta_B^2$
at the resonance,
where the prime denotes the derivative with respect to the distance
along the propagation direction. In the following we will estimate this
derivative by the coherence length, i.e. $^\prime\rightarrow1/\lambda$.

One has $\Delta_a=\Delta_{\rm pl}$ at the electron density
\begin{equation}\label{eq:ne_msw1}
  n_e\sim7.1\times10^8\,m_{\mu{\rm eV}}^2\,{\rm cm}^{-3}\,.
\end{equation}
This leads to an adiabatic resonance at energies satisfying
\begin{eqnarray}\label{eq:msw1}
  E&\gtrsim&20\,g_{11}^{-2}m_{\mu{\rm eV}}^2\lambda_{\rm pc}^{-1}
  B_{\rm mG}^{-2}\,{\rm TeV}\,,\nonumber\\
  E&\lesssim&4.4\,m_{\mu{\rm eV}}B_{\rm mG}^{-1}\,{\rm TeV}\,.
\end{eqnarray}
In Eq.~(\ref{eq:msw1}), the first condition results from the adiabaticity
requirement with $\Delta_{\rm osc}^\prime\sim\Delta_{\rm pl}/\lambda$
and the second from $\Delta_{\rm CM}\lesssim\Delta_a$. As a result,
strong $\gamma-$ray flux suppression is expected at energies
satisfying Eq.~(\ref{eq:msw1}).

One has $\Delta_{\rm CM}=-\Delta_{\rm pl}$ at the energy
\begin{equation}\label{eq:msw2}
  E\sim5.2\,\left(\frac{n_e}{10^3\,{\rm cm}^{-3}}\right)^{1/2}
  B_{\rm mG}^{-1}\,{\rm GeV}\,,
\end{equation}
This leads to an adiabatic resonance provided that at this energy
$\Delta_a\lesssim|\Delta_{\rm pl}|$ and that the adiabaticity condition
is fulfilled with $\Delta_{\rm osc}^\prime\sim\Delta_{\rm CM}/\lambda$.
This translates into
\begin{eqnarray}\label{eq:cond_msw2}
  n_e&\gtrsim&7.1\times10^8\,m_{\mu{\rm eV}}^2\,{\rm cm}^{-3}\,,\nonumber\\
  E&\lesssim&0.95\,g_{11}^2\lambda_{\rm pc}\,{\rm TeV}\,,
\end{eqnarray}
respectively. Note that the first condition in Eq.~(\ref{eq:cond_msw2})
together with Eq.~(\ref{eq:msw2}) implies
$E\gtrsim4.4\,m_{\mu{\rm eV}}B_{\rm mG}^{-1}\,{\rm TeV}$, such that
this type of resonance starts to be relevant at energies just above
where the first type of resonance ceases. This is understood since
for the first resonance we had $\left|\Delta_{\rm CM}\right|\lesssim\Delta_a$,
whereas for the second $\Delta_a\lesssim\Delta_{\rm pl}=\left|\Delta_{\rm CM}\right|$.

Eq.~(\ref{eq:msw1}) and the second inequality in Eq.~(\ref{eq:cond_msw2})
combined thus lead to resonances in the energy range
\begin{eqnarray}\label{eq:msw}
  E&\gtrsim&20\,g_{11}^{-2}m_{\mu{\rm eV}}^2\lambda_{\rm pc}^{-1}
  B_{\rm mG}^{-2}\,{\rm TeV}\,,\nonumber\\
  E&\lesssim&0.95\,g_{11}^2\lambda_{\rm pc}\,{\rm TeV}\,,
\end{eqnarray}
provided the following conditions are satisfied:
\begin{eqnarray}\label{eq:cond_msw}
  g_{11}&\gtrsim&2.1
  \left(\frac{m_{\mu{\rm eV}}}{\lambda_{\rm pc}B_{\rm mG}}\right)^{1/2}
  \,,\nonumber\\
  n_e&\gtrsim&7.1\times10^8\,m_{\mu{\rm eV}}^2\,{\rm cm}^{-3}\,.
\end{eqnarray}
Here the first condition comes from the requirement that Eq.~(\ref{eq:msw})
represents a finite energy range, and the second condition results
from Eq.~(\ref{eq:ne_msw1}) and the first inequality in
Eq.~(\ref{eq:cond_msw2}). For a given coupling $g_{11}$ resonances
occur only for coherence lengths (and thus system sizes) satisfying
\begin{equation}\label{eq:lambda_res}
  d\gtrsim\lambda\gtrsim1.4\times10^{19}\,g_{11}^{-2}m_{\mu{\rm eV}}
  B_{\rm mG}^{-1}\,{\rm cm}\,.
\end{equation}

Inserting Eq.~(\ref{eq:lambda_res}) into Eq.~(\ref{eq:msw}) thus
also leads to the inequalities
\begin{equation}\label{eq:maxmin_res}
  E^{\rm r}_{\rm min}\lesssim4.4\,m_{\mu{\rm eV}}\,B_{\rm mG}^{-1}\,{\rm TeV}
  \lesssim E^{\rm r}_{\rm max}\,.
\end{equation}
for the minimal and maximal photon energy $E^{\rm r}_{\rm min}$
and $E^{\rm r}_{\rm max}$, respectively, at which resonances occur.

\subsection{Non-resonant Oscillations}
Let us denote the coherence length of the magnetic field by $\lambda$
and the propagation length with $d$, so that we have
$N\sim d/\lambda\gtrsim1$ domains. For
$\Delta_{\rm osc}\lambda\lesssim1$, we have from
Eq.~(\ref{eq:p_0}) $P_0\sim(\Delta_B \lambda)^2$. In the
opposite limit, $\Delta_{\rm osc}\lambda\gtrsim1$, we have
$P_0\sim2(\Delta_B/\Delta_{\rm osc})^2$.
Altogether, with Eq.~(\ref{eq:p}) this yields for the photon
survival probability $P_{\gamma\to\gamma}\equiv1-P_{\gamma\to a}$,
\begin{equation}\label{eq:p2}
  P_{\gamma\to \gamma}\sim\frac{1}{3}\left[2+\exp\left(-\,\frac{3}{2}\,\,{\rm min}
  \left[\Delta_B^2 d\lambda,2N(\Delta_B/\Delta_{\rm osc})^2\right]
  \right)\right]\,.
\end{equation}
For given axion parameters, the spectra predicted within scenarios
of $\gamma-$ray sources will be modified by this factor.
Eq.~(\ref{eq:p2}) can only be of order unity and thus lead to observable
effects, if {\it both} $\Delta_B^2 d\lambda\gtrsim1$
and $2(\Delta_B/\Delta_{\rm osc})^2N\gtrsim1$.
With Eq.~(\ref{eq:frequencies}), the first condition yields
\begin{equation}\label{eq:condition1}
  \left(d\lambda\right)^{1/2}\gtrsim
  5.7\times10^{19}\,g_{11}^{-1}\,B_{\rm mG}^{-1}\,{\rm cm}\,.
\end{equation}
This results in the two conditions
\begin{equation}\label{eq:condition1a}
  d\gtrsim5.7\times10^{19}\,g_{11}^{-1}\,B_{\rm mG}^{-1}\,{\rm cm}\,,
\end{equation}
and
\begin{equation}\label{eq:condition1b}
  1\le N^{1/2}\equiv\left(\frac{d}{\lambda}\right)^{1/2}
  \lesssim0.054\,d_{\rm pc}g_{11}B_{\rm mG}\,,
\end{equation}
or equivalently
\begin{equation}\label{eq:condition1c}
  \lambda=\frac{d}{N}\gtrsim320\,g_{11}^{-2}d_{\rm pc}^{-1}
  B_{\rm mG}^{-2}\,{\rm pc}\,,
\end{equation}
where $d_{\rm pc}\equiv(d/{\rm pc})$.

The second condition from Eq.~(\ref{eq:p2}) leads to the three separate 
inequalities $\Delta_B\gtrsim\Delta_{\rm a}N^{-1/2}$,
$\Delta_B\gtrsim\left|\Delta_{\rm pl}\right|N^{-1/2}$, and
$\Delta_B\gtrsim\Delta_{\rm CM}N^{-1/2}$. Substituting
Eq.~(\ref{eq:frequencies}) results in
\begin{eqnarray}\label{eq:condition2}
	E&\gtrsim&15\,g_{11}^{-1}\,m_{\mu{\rm eV}}^2\,N^{-1/2}\,B_{\rm mG}^{-1}
	\,{\rm TeV}\,,\nonumber\\
  E&\gtrsim&21\left(\frac{n_e}{10^3\,{\rm cm}^{-3}}\right)\,
  g_{11}^{-1}\,N^{-1/2}\,B_{\rm mG}^{-1}\,{\rm MeV}\,,\\
	E&\lesssim&1.3\,g_{11}\,N^{1/2}\,B_{\rm mG}^{-1}\,{\rm TeV}\,.\nonumber
\end{eqnarray}
Note that the scaling of these energies with $\lambda$ are different
from the scaling of the resonance energies Eq.~(\ref{eq:msw}): Large
coherence lengths, or small $N$, favor a broad resonance energy range,
whereas small coherence lengths will tend to lead to non-resonant
transitions, as long as Eq.~(\ref{eq:condition1b}) is satisfied.
The absolute minimal and maximal photon energies at which non-resonant
oscillations occur are given by setting $N=1$ in Eq.~(\ref{eq:condition2}).

We can now eliminate $N$ or equivalently $\lambda$ from
Eq.~(\ref{eq:condition2}) by using Eq.~(\ref{eq:condition1a}):
\begin{eqnarray}\label{eq:condition2a}
	E&\gtrsim&278\,g_{11}^{-2}\,m_{\mu{\rm eV}}^2\,d_{\rm pc}^{-1}
	\,B_{\rm mG}^{-2}\,{\rm TeV}\,,\nonumber\\
  E&\gtrsim&389\left(\frac{n_e}{10^3\,{\rm cm}^{-3}}\right)\,
  g_{11}^{-2}\,d_{\rm pc}^{-1}\,B_{\rm mG}^{-2}\,{\rm MeV}\,,\\
	E&\lesssim&0.07\,g_{11}^2\,d_{\rm pc}\,{\rm TeV}\,.\nonumber
\end{eqnarray}

For Eqs.~(\ref{eq:condition2}) and~(\ref{eq:condition2a}) to
be satisfied for a finite energy range the following
conditions are implied:
\begin{eqnarray}\label{eq:conditions3}
  g_{11}&\gtrsim&3.4\,m_{\mu{\rm eV}}\,N^{-1/2}\gtrsim0.13
  \left(\frac{m_{\mu{\rm eV}}}{d_{\rm pc}B_{\rm mG}}\right)^{1/2}\,,\\
  n_e&\lesssim&6.2\times10^7\,g_{11}^2\,N
  \,{\rm cm}^{-3}\lesssim1.8\times10^5\,g_{11}^4d_{\rm pc}^2B_{\rm mG}^2
  \,{\rm cm}^{-3}\nonumber\,,
\end{eqnarray}
where in the second expressions we have substituted
Eq.~(\ref{eq:condition1b}). The condition on $g_{11}$ is very
similar to the resonance condition Eq.~(\ref{eq:cond_msw}).

Coupling and mass of the QCD axion roughly satisfy
$g_{11}\sim10^{-5}m_{\mu{\rm eV}}$, and thus experimental
limits imply $m_a\lesssim1\,$eV. The first condition
in Eq.~(\ref{eq:conditions3}) then implies
\begin{eqnarray}\label{eq:qcd}
  N&\gtrsim&1.2\times10^{11}\,,\\
  d_{\rm pc}B_{\rm mG}&\gtrsim&18\,g_{11}^{-1}
  \sim1.8\times10^6m_{\mu{\rm eV}}^{-1}\,,\nonumber
\end{eqnarray}
where in the second line we have used Eq.~(\ref{eq:condition1a}).
In 2005, the PVLAS~\cite{Zavattini:2005tm} experiment has seen indications
for axion-photon mixing with
\begin{eqnarray}\label{eq:pvlas}
  m_a&\simeq&1.3\,{\rm meV}\nonumber\\
  g_{\gamma a}&\simeq&3\times10^{-6}\,{\rm GeV}^{-1}\,,
\end{eqnarray}
which since recently, however, is considered to probably have been
an experimental effect~\cite{Zavattini:2007ee}. In any case, the "PVLAS axion"
would fulfill the first condition in Eq.~(\ref{eq:conditions3}).

\subsection{Other Conditions}
A further condition comes from the requirement that the length
scale over which the emission is created
cannot be larger than the variability time scale of the source.
The size of the emission region is determined by the length
scale over which the plasma becomes transparent to $\gamma\gamma$
pair production.
Emissions of active and radio galaxies often vary on times scales
of days or even hours~\cite{m87-variability,Krennrich:2002as},
corresponding to scales $\lesssim10^{14-15}\,$cm. If these are
the same length scales over which significant photon-axion
conversion occurs, then one also has the condition
$\lambda\le d\lesssim10^{14-15}\,$cm, except in case of relativistic
beaming.

\section{Prospects for specific Sources}
\label{sec:sources}
A lower limit on the source magnetic field times the propagation
length can actually be obtained by requiring that it accelerates
cosmic rays of charge $Ze$ up to the maximal energy $E_{\rm cr}$
observed from that source:
\begin{equation}\label{eq:hillas}
  d\gtrsim\frac{E_{\rm cr}}{ZeB}\simeq3.3\times10^{12}\,Z^{-1}
  \left(\frac{E_{\rm cr}}{\rm TeV}\right)\,B_{\rm mG}^{-1}
  \,{\rm cm}\,.
\end{equation}
This implies that sources that accelerate particles up to
\begin{equation}
  E_{\rm cr}\gtrsim1.7\times10^{19}\,Z g_{11}^{-1}\,{\rm eV}
\end{equation}
fulfill the condition Eq.~(\ref{eq:condition1a}) independently
of the magnetic field strength. The coherence length of the fields depends
on the acceleration model and has to satisfy the additional condition
Eq.~(\ref{eq:condition1c}). Sources
of ultra-high energy cosmic rays with energies $E_{\rm cr}>10^{18}\,$eV
are thus promising objects that may exhibit axion-photon mixing induced
modifications of their photon spectra for couplings
$g_{11}\gtrsim1.7\,(Z\,10^{19}\,{\rm eV}/E_{\rm cr})$.

Whereas this condition for the occurrence of photon-axion oscillations
does not depend on size or magnetic field strength of the accelerator,
the energies at which such oscillations could modify photon spectra do
depend on these parameters: Eqs.~(\ref{eq:msw}) and~(\ref{eq:condition2a})
show that the minimal and maximal energies at which significant oscillation
effects can occur, are given by
\begin{eqnarray}\label{eq:maxmin}
  E_{\rm min}&\sim&19\,g_{11}^{-2}\,m_{\mu{\rm eV}}^2\,B_{\rm mG}^{-1}
  \left(\frac{Z\,10^{18}\,{\rm eV}}{E_{\rm cr}}\right)\,{\rm TeV}\nonumber\\
  &\sim&18\,g_{11}^{-2}\,m_{\mu{\rm eV}}^2\,d_{\rm pc}\,
  \left(\frac{Z\,10^{18}\,{\rm eV}}{E_{\rm cr}}\right)^2\,{\rm TeV}\nonumber\\
  E_{\rm max}&\sim&0.95\,g_{11}^2\,d_{\rm pc}\,{\rm TeV}\,,
\end{eqnarray}
where we have used the relation Eq.~(\ref{eq:hillas}).

Among the extragalactic objects which have been seen in very high
energy $\gamma-$rays and where photon-axion conversion could occur
are Markarian 421~\cite{Konopelko:2003zr}, Markarian
501~\cite{mrk501,Konopelko:2003zr},
the blazar 1ES 1101-232~\cite{Aharonian:2007nq}, and the
variable core~\cite{m87-variability} and flaring knots~\cite{Cheung:2007wp}
of M87. Markarian 421 is also variable in $\gamma-$rays\cite{Krennrich:2002as}.
Recent reviews on leptonic and hadronic models of blazar emission can
be found in Refs.~\cite{Torres:2003zb,Boettcher:2006pd}. The spectrum of
a typical AGN has a double-peaked power spectrum. In leptonic models, the
high energy peak is caused by inverse Compton scattering of accelerated
electrons on the ambient photon field, whereas the low energy peak is
due to synchrotron emission of these same electrons. The energies and
relative power flux in these peaks thus contains information on the
magnetic fields in the emission region. The estimated magnetic field strengths
range between $\sim 10$ mG and $\sim 10$ G over scales, depending on the models used,
typically of order $d\gtrsim10^{-3}\,$pc. This would also suggest possible
cosmic ray acceleration up to $\sim10^{18}\,$eV.
For axion parameters $g_{11}\sim1$, $m_{\mu{\rm eV}}\sim1$,
Eq.~(\ref{eq:maxmin}) then suggests photon-axion oscillation effects
at GeV energies, consistent with the findings of Ref.~\cite{Hooper:2007bq}.

For the galactic center fields of order 10 G and up to $10^4\,$G
have been discussed in Ref.~\cite{Aharonian:2005ti,Aharonian:2004jr}.
The galactic center is sometimes thought to accelerate cosmic rays
up to $\sim10^{18}\,$eV~\cite{Anchordoqui:2003vc}. Eq.~(\ref{eq:maxmin})
would then imply significant non-resonant axion-photon oscillations down
to MeV energies.

\subsection{Former Work}
The original indications from PVLAS for axion-like particles with
rather strong couplings, Eq.~(\ref{eq:pvlas}), although in contradiction
with present astrophysical constraints,
has motivated the study of several possible astrophysical effects.
Ref.~\cite{Dupays:2006dg,Dupays:2006hz}, considered variable $\gamma-$ray
light curves from double pulsars as signatures for
the PVLAS axion-photon mixing parameters.
In this scenario, a surface pulsar magnetic field $B\sim10^{12}\,$G
yields $B\sim10^4\,$G, $N\sim1$ in the accretion region between the
pulsars where the $\gamma-$rays are produced, and the Goldreich-Julian plasma density is $n_e\sim10^3\,{\rm cm}^{-3}$. 
Eq.~(\ref{eq:condition2}) implies a modulation for
${\rm MeV}\lesssim E\lesssim100\,$GeV, as
obtained in Ref.~\cite{Dupays:2006dg,Dupays:2006hz}. Note
that for less coherent fields between the two compact stars,
$N>1$, the modulation would extend down to lower energies.

Ref.~\cite{Mirizzi:2007hr} considered the modification of the galactic
center $\gamma-$ray flux by PVLAS axions, for $B\sim\mu$G,
$\lambda\sim0.01\,$pc, $d\sim10\,$kpc, thus $N\sim10^6$.
Thus Eq.~(\ref{eq:condition2}) results in a modification at
energies $10\,{\rm TeV}\lesssim E\lesssim10^{12}\,$TeV. Note that
the coherence length of the Galactic magnetic field is not very
well known~\cite{vallee}. For $\lambda\gtrsim$ pc,
the effect would disappear for all observable energies
$E\lesssim100\,$TeV. Measurement of Faraday rotation from
pulsar pairs suggest that the cell size of the random component of
the galactic magnetic field may indeed be of order $50\,$pc~\cite{ohno,han}.

No resonances are expected in these two scenarios due to the
second condition in Eq.~(\ref{eq:cond_msw}), unless
$m\lesssim10^{-8}\,$eV.

Ref.~\cite{Mortsell:2003ya} constrained axion-photon mixing
by using quasar spectra mostly in the optical, considering
mixing in intergalactic fields of strength $B\lesssim10^{-9}\,$G
only. In this case such spectra are only sensitive
to axion masses $m_a\lesssim10^{-15}\,$eV, consistent with
Eqs.~(\ref{eq:msw}) and~(\ref{eq:condition2a}). For larger
axion masses $m\sim10^{-10}\,$eV, such conversions in intergalactic
magnetic fields would show up at TeV energies, as discussed in
Ref.~\cite{De Angelis:2007yu}.

Ref.~\cite{Hooper:2007bq} discussed the modification of $\gamma-$ray spectra
from AGNs by axion-photon mixing, using magnetic fields and length
scales according to the Hillas criterion for accelerating particles
to the pertinent energies. However, they did not consider resonances.

Note that emission from the surface of a neutron star or magnetar
with $d\sim10^6\,$cm, $B\sim10^{12}$--$10^{15}\,$G,
$n_e\sim10^{34}\,{\rm cm}^{-3}$ according to the second condition
in Eq.~(\ref{eq:conditions3}) could not give significant non-resonant
transitions for $g_{11}\lesssim10^5$. Furthermore, according to
Eq.~(\ref{eq:msw}), resonances can only occur at very low energies
$E\lesssim0.3\,g_{11}^2\,$eV.

\subsection{The central engine of AGNs}

Black hole accretion launches jets by magnetohydrodynamical processes.
For spherical Bondi-accretion, the accretion rate is
$\dot M\sim4\pi r^2n_em_p\beta$ where $m_p$ is the proton mass
and $\beta\sim\beta_0(r_{\rm S}/r)^{1/2}$ are the density and velocity of the
accretion flow, respectively. Here, $\beta_0\sim0.1$ and
$r_{\rm S}=2G_{\rm N}M\simeq2.96\times10^{14}M_9\,{\rm cm}$
is the Schwarzschild radius for a black hole of mass $M=M_9\,M_\odot$.
Introducing the Eddington luminosity,
$L_{\rm Edd}(M)\simeq1.3\times10^{47}M_9\,{\rm erg}\,{\rm s}^{-1}$,
the accretion rate can be written as $\dot M=f_{\rm Edd}L_{\rm Edd}(M)/\eta$,
where $\eta$ is the efficiency with which accretion is converted into electromagnetic
radiation, and $f_{\rm Edd}=L_{\rm bol}/L_{\rm Edd}(M)$ is the
ratio of the bolometric luminosity to the Eddington luminosity.
AGNs have duty cycles of a few percent and in their active periods
have $\eta\sim f_{\rm Edd}\sim0.1$.

From this we can estimate the plasma density in the accretion flow by
$n_e\sim\dot M/(4\pi r^2 m_p\beta)$ which close to the Schwarzschild
radius yields
\begin{equation}\label{eq:n_e_agn}
  n_e\sim2.6\times10^9\,\left(\frac{f_{\rm Edd}}{\eta\beta_0}\right)
  M_9^{-1}\,\left(\frac{r}{r_{\rm S}}\right)^{-3/2}\,{\rm cm}^{-3}\,.
\end{equation}

If the magnetic fields are roughly in equipartition with the accretion
flow at radius $r$, one has $B^2/(8\pi)\sim n_em_p\beta^2$.
This results in~\cite{Camenzind:2004ae}
\begin{eqnarray}\label{eq:B_agn}
  B&\sim&3.9\,\beta_0^{1/2}
  \left(\frac{\dot M}{M_\odot\,{\rm yr}^{-1}}\right)^{1/2}
  \,M_9^{-1}\,
  \left(\frac{r}{r_{\rm S}}\right)^{-5/4}\,{\rm kG}\nonumber\\
  &=&5.9\,\left(\frac{\beta_0 f_{\rm Edd}}{\eta}\right)^{1/2}
  \,M_9^{-1/2}\,
  \left(\frac{r}{r_{\rm S}}\right)^{-5/4}\,{\rm kG}\,.
\end{eqnarray}
where in the second expression the accretion rate has been written
in terms of the Eddington luminosity.
Note that these are very rough order of magnitude estimates
which, however, will be sufficient for our purpose.

If accretion is non-spherical, but instead occurs in a disk, as is
in general the case for very luminous AGNs, for a given accretion
rate, the plasma density and magnetic field strengths tend to be larger
than the estimates Eqs.~(\ref{eq:n_e_agn}) and~(\ref{eq:B_agn}).
In these and all subsequent formulas, this is essentially mimicked
by an electromagnetic efficiency $\eta$ smaller by a factor
of roughly the height of the disk relative to its radius. This will
be very roughly the Lorentz factor of the resulting jet. The geometry is,
however, more complicated in this case and photon-axion oscillations
may be suppressed perpendicular to the disk. The strongest effects
may then only occur in directions of the plane of the disk. As we will
see below, both the minimal Eddington ratio $f_{\rm Edd}$ and the energy
for which oscillation effects occur are proportional to $\eta$, whereas
other parameters are largely insensitive to $\eta$.

In any case, more detailed theoretical models of magnetized accretion
disks lead to magnetic fields of order $10^4\,$G in the disk and
$\sim10^2\,$G above the disk~\cite{magnetized-disk}. Such values are consistent
with the rough estimates in Eq.~(\ref{eq:B_agn}). In addition, supermassive
black holes may have a dipole magnetic field with a maximal field strength
of $\sim2\times10^{10}\,M_9^{-1}\,$G at the Schwarzschild
radius~\cite{kardashev}.

But how coherent are these fields? We note that they may be made
coherent by dynamo effects and/or magnetorotational instability (MRI)
Jet launching by MHD effects may indeed require coherent, very
non-thermal fields. In this case the fields would be coherent on
roughly the Schwarzschild radius.

Magnetic fields of order 10 G in the vicinity of the radio galaxy
M87 have been discussed in Ref.~\cite{Neronov:2007vy}. This value
is consistent with the estimate Eq.~(\ref{eq:B_agn}) if M87 is
in a low accretion state, $f_{\rm Edd}\sim10^{-6}$, which is the case
according to present measurements and theories~\cite{Neronov:2007vy}.

The variability constraint is fulfilled automatically since AGNs
cannot be variable on time scales smaller than the Schwarzschild
radius.

\subsubsection{Resonances}
Comparing the second condition in Eq.~(\ref{eq:cond_msw}) with
Eq.~(\ref{eq:n_e_agn}) implies the condition
\begin{equation}\label{eq:cond_smbh_res1}
  M\lesssim 3.7\times10^9\,m_{\mu{\rm eV}}^2
  \left(\frac{f_{\rm Edd}}{\eta\beta_0}\right)\,M_\odot
\end{equation}
for a resonance to occur above the Schwarzschild radius.

\begin{figure}[ht]
\includegraphics[width=0.5\textwidth,clip=true]{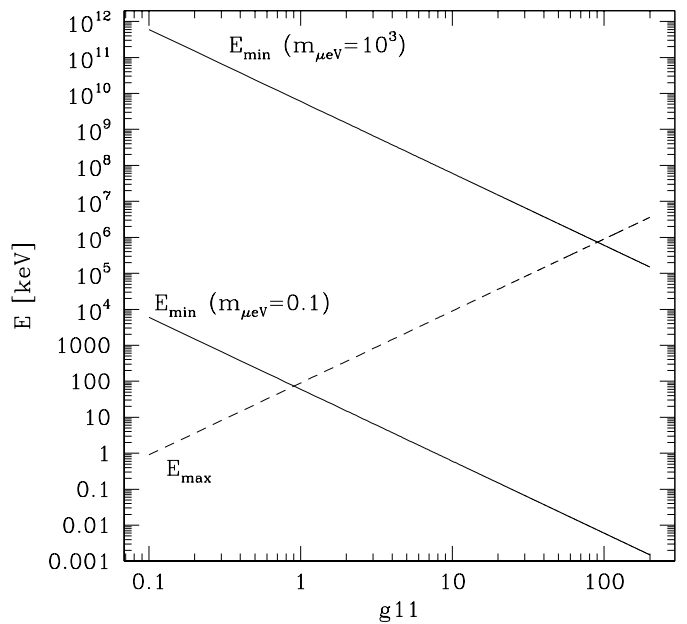}
\caption{Maximal and minimal energies Eq.~(\ref{eq:e_agn_res})
for which axion-photon resonances occur, for $M\sim10^6\,M_\odot$,
$f_{\rm Edd}\sim10^{-3}$, $\beta_0=\eta=0.1$, $\lambda=r_{\rm S}$.
Whenever $E_{\rm min}\le E_{\rm max}$, adiabatic resonances occur.
Note that in this case also the conditions
Eqs.~(\ref{eq:cond_smbh_res1}), (\ref{eq:cond_smbh_res2}),
and~(\ref{eq:coh_agn}) are fulfilled.}
\label{fig1}
\end{figure}

Using Eq.~(\ref{eq:B_agn}) and assuming $\lambda\propto r$,
Eq.~(\ref{eq:msw}) shows that the widest energy range is
achieved for the smallest radii. Further, the first
condition in Eq.~(\ref{eq:cond_msw}) implies the least
stringent condition on $g_{11}$ for the smallest radii.
We thus get
\begin{equation}\label{eq:cond_smbh_res2}
  g_{11}\gtrsim8.8\times10^{-2}\,N^{1/2}\,m_{\mu{\rm eV}}^{1/2}\,
  M_9^{-1/4}\left(\frac{\eta}{\beta_0 f_{\rm Edd}}\right)^{1/4}\,.
\end{equation}
Thus, when the above conditions Eq.~(\ref{eq:cond_smbh_res1}) and
Eq.~(\ref{eq:cond_smbh_res2}) are fulfilled, a strong $\gamma-$ray
flux suppression occurs for an energy interval given by substituting
the field strength Eq.~(\ref{eq:B_agn}) close to the Schwarzschild
radius into Eq.~(\ref{eq:msw}),
\begin{eqnarray}\label{eq:e_agn_res}
  E&\gtrsim&6\times 10^3\,g_{11}^{-2}\,m_{\mu{\rm eV}}^2\,N
  \,\left(\frac{\eta}{\beta_0f_{\rm Edd}}\right)\,{\rm eV}\,,\nonumber\\
  E&\lesssim&90\,g_{11}^2\,N^{-1}\,M_9\,{\rm MeV}\,,
\end{eqnarray}
where $N\equiv\lambda/r_{\rm S}$ is the number of coherent domains
within a Schwarzschild range. 
Note that the lower limit does not depend on the black hole mass.
In Fig.~\ref{fig1} we show the energy range Eq.~(\ref{eq:e_agn_res})
for fields coherent over the Schwarzschild radius, $N=1$, for an
AGN of the type of Mrk421 or Mrk501, with $M\sim10^6\,M_\odot$,
$f_{\rm Edd}\sim10^{-3}$. For couplings approaching the experimental
upper limit, $g_{a\gamma}\lesssim10^{-10}\,{\rm GeV}^{-1}$, resonances
can occur down to energies
$\sim600\,(m/0.1\mu{\rm eV})^2(f_{\rm Edd}/10^{-3})^{-1}\,$eV
which can extend into the keV regime.

The condition on the coherence scale Eq.~(\ref{eq:lambda_res}) reads
\begin{eqnarray}\label{eq:coh_agn}
  f_{\rm Edd}&\gtrsim&6.1\times10^{-5}\,\left(\frac{\eta}{\beta_0}\right)
  g_{11}^{-4}\,m_{\mu{\rm eV}}^2\,M_9^{-1}\,,\\
  N^{-1}&\equiv&\frac{\lambda}{r_{\rm S}}\gtrsim7.8\times10^{-3}\,
  g_{11}^{-2}m_{\mu{\rm eV}}M_9^{-1/2}
  \left(\frac{\eta}{\beta_0 f_{\rm Edd}}\right)^{1/2}\,,\nonumber
\end{eqnarray}
where the first condition follows from the second observing that
$N\ge1$.

\subsubsection{Non-resonant Oscillations}
Substituting $d=r=r_{\rm S}$ in the conditions Eq.~(\ref{eq:condition1a})
and Eq.~(\ref{eq:condition1b}) and in Eq.~(\ref{eq:B_agn}) results in
requirements similar to Eq.~(\ref{eq:coh_agn}),
\begin{eqnarray}\label{eq:cond1_smbh}
  f_{\rm Edd}&\gtrsim&1.1\times10^{-3}\,\left(\frac{\eta}{\beta_0}\right)
  g_{11}^{-2}\,M_9^{-1}\,,\\
  N^{-1}&\equiv&\frac{\lambda}{r_{\rm S}}\gtrsim1.1\times10^{-3}\,
  g_{11}^{-2}M_9^{-1}\frac{\eta}{\beta_0 f_{\rm Edd}}\,.\nonumber
\end{eqnarray}
We remark that the first constraint is easily fulfilled for luminous
AGNs which have $f_{\rm Edd}\gtrsim10^{-2}$, especially for relatively
large axion-photon coupling. The central black hole of our Galaxy has
$f_{\rm Edd}\sim10^{-8}$.

Note that the condition on the coherence length in Eq.~(\ref{eq:cond1_smbh})
tends to be less stringent than Eq.~(\ref{eq:coh_agn}), consistent with
the fact that more coherent fields tend to lead to resonances.

\begin{figure}[ht]
\includegraphics[width=0.5\textwidth,clip=true]{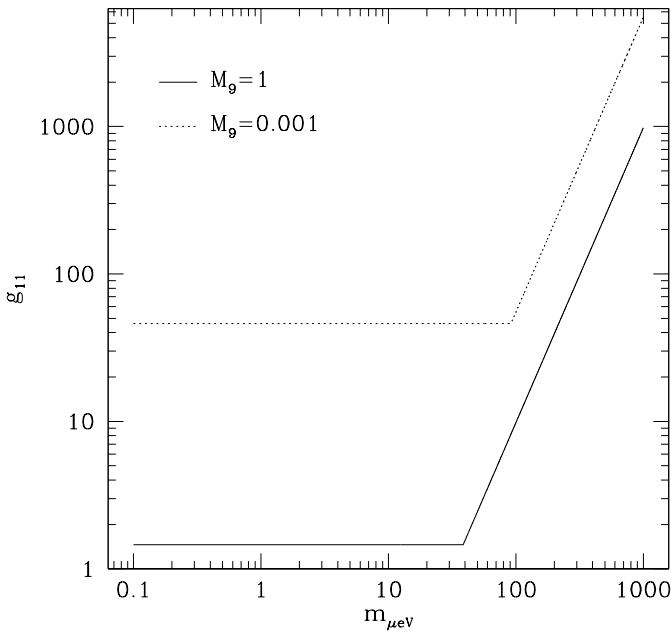}
\caption{Range of coupling constants and axion masses
Eq.~(\ref{eq:cond3_smbh}) for which non-resonant axion-photon
oscillations can influence the $\gamma-$ray spectra from
AGNs with $f_{\rm Edd}=10^{-3}$.}
\label{fig2}
\end{figure}

Using the plasma density Eq.~(\ref{eq:n_e_agn}) and again
substituting $d=r=r_{\rm S}$ and Eq.~(\ref{eq:B_agn}),
the two conditions in Eq.~(\ref{eq:conditions3}) translate into
\begin{eqnarray}\label{eq:cond3_smbh}
  g_{11}&\gtrsim&5.5\times10^{-3}\,m_{\mu{\rm eV}}^{1/2}M_9^{-1/4}
  \left(\frac{\eta}{\beta_0 f_{\rm Edd}}\right)^{1/4}\,,\nonumber\\
  g_{11}&\gtrsim&\,0.46\beta_0^{-1/2}M_9^{-1/2}\,.
\end{eqnarray}
This implies that emission from massive black holes can be sensitive
to rather small coupling constants. The range of coupling constants
given by Eq.~(\ref{eq:cond3_smbh}) is shown in Fig.~\ref{fig2}.

Thus, when the above conditions Eqs.~(\ref{eq:cond1_smbh}) and
Eqs.~(\ref{eq:cond3_smbh}) are fulfilled,
a $\gamma-$ray flux suppression by about a factor $\simeq2/3$
occurs for energies
\begin{eqnarray}\label{eq:e_agn}
  E&\gtrsim&2.5\,g_{11}^{-1}\,m_{\mu{\rm eV}}^2\,
  N^{-1/2}\left(\frac{\eta}{\beta_0 f_{\rm Edd}}\right)^{1/2}
  M_9^{1/2}\,{\rm MeV}\nonumber\\
  &&\gtrsim0.083\,g_{11}^{-2}\,m_{\mu{\rm eV}}^2
  \left(\frac{\eta}{\beta_0 f_{\rm Edd}}\right)\,{\rm MeV}\,,\\
  E&\lesssim&0.22\,g_{11}^{-1}\,
  N^{1/2}\,\left(\frac{\eta}{\beta_0 f_{\rm Edd}}\right)^{1/2}
  \,M_9^{1/2}\,{\rm MeV}\nonumber\\
  &&\lesssim6.6\,M_9\,{\rm MeV}\,,\nonumber
\end{eqnarray}
where for the second expressions we have used the constraint on the
number of domains in the second line of Eq.~(\ref{eq:cond1_smbh}). The third
condition that can be obtained from Eq.~(\ref{eq:condition2}) is
less stringent than the one given.

Note that the energies Eq.~(\ref{eq:e_agn}) tend to be in the hard
X-ray to soft $\gamma-$ray range except for very strong couplings
of the PVLAS type for which these energies can extend down to the
optical range. Such photons are not absorbed by pair production
and there is plenty of data at such energies that could
be searched for the spectral features discussed here.

\subsection{AGN jets and hot-spots}
AGN jets and hot-spots seem to be likely candidates to fulfill the requirements for observable photon-axion conversion, as the condition Eq.~(\ref{eq:condition1a}) is
satisfied for all couplings $g_{11}\gtrsim10^{-2}$
in such environments since even the transverse
dimension of the jets are of order kpc and fields are at least
of order $100\,\mu$G~\cite{sikora,sreekumar,ostrowski}. The first
condition in Eq.~(\ref{eq:conditions3}) shows that this allows
significant effects down to couplings
$g_{11}\gtrsim0.013\,m_{\mu{\rm eV}}^{1/2}$, provided the coherence
scale $\lambda\sim$ pc. Furthermore, the plasma density in jets is assumed as 
$n_e\lesssim10^5\,{\rm cm}^{-3}$, about a factor 100 higher than
the ambient density of an average galaxy, as expected for termination
shocks. Thus, the second condition in Eq.~(\ref{eq:conditions3}) is
also satisfied, unless $g_{11}\lesssim0.1$. According to
Eq.~(\ref{eq:condition2a}), the resulting features would show up
at TeV energies. However, due to Eq.~(\ref{eq:cond_msw2}), resonances
would only occur for $m_a\lesssim1.2\times10^{-8}\,$eV.

In the proton synchrotron model, field strengths of up to $10\,$mG
have been discussed in Ref.~\cite{aharonian}. For axions with
$g_{11}\sim1$, $m_a\sim\mu$eV
Eqs.~(\ref{eq:condition2}) and~(\ref{eq:condition2a}) thus imply
possible effects down to $\sim$ GeV energies, if $\lambda\sim0.1\,$pc.

\begin{figure}[ht]
\includegraphics[width=0.5\textwidth,clip=true]{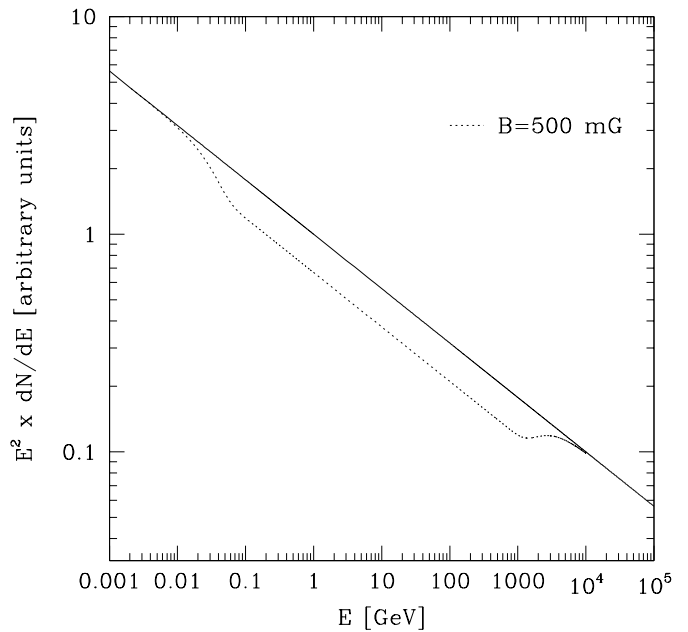}
\caption{Modification of the $\gamma-$ray spectrum of an AGN
of mass $M=10^9\,M_\odot$, for $B=0.5\,$G over $2\times10^6$
Schwarzschild radii, or $\sim200\,$pc
with coherence length $\lambda$ equal to 10 times the
Schwarzschild radius, or $\sim10^{-3}\,$pc.
The assumed injection spectrum is $\propto E^{-2.25}$.
The axion parameters are $g_{\gamma a}=10^{-11}\,{\rm GeV}^{-1}$, $m_a=1\mu$eV.
The modification factor $P_{\gamma\to\gamma}$ is given
by Eq.~(\ref{eq:p2}). Spectral modification by pair production has
not been taken into account and is negligible for redshifts
$z\lesssim0.03$~\cite{Stecker:2006vx}.}
\label{fig3}\end{figure}

As an example for the effect of non-resonant oscillations
we show in Fig.~\ref{fig3} the modification of the $\gamma-$ray
spectrum of a quasar around the maximal energy given by
Eq.~(\ref{eq:condition2}).
The parameters used are typical for AGNs with central black
hole masses around $10^9\,M_\odot$. The magnetic field is assumed
of order a Gauss over length scales $10^6$ times the Schwarzschild
radius. Such fields can occur in models of kpc scale jets which emit MeV
$\gamma-$rays produced by inverse Compton scattering of accelerated
electrons on low energy synchrotron photons and external
photons~\cite{sikora,Boettcher:2006pd}, and in ultra-compact
jets~\cite{lobanov}. However, variability over scales of
months or less would imply that the length scales over which
significant photon-axion conversion occurs would be much larger
than the size of the region where most of the emission is produced.
This region would have to be close to the black hole, except if the jet
is oriented toward the observer such that relativistic beaming compresses
the time scales.

\section{Conclusions}
\label{sec:conclusions}
We have investigated the possibilities for successful photon-axion conversion in $\gamma$-ray sources such as the discs and jets of AGNs. We have derived conditions for resonant and non-resonant oscillations and find that a significant conversion is possible for standard axion parameters with coupling $g_{11}\sim 1$ and axion mass $m_{\mu{\rm eV}}\sim 1$, which are allowed by present experimental and astrophysical constraints. Of course the efficiency of conversion depends on the strength of the magnetic field in the vicinity of the black hole. Values between $\sim$ 0.01 G and several G are under discussion. Resonant effects, leading to gaps in the observed spectrum, are observable for plasma densities $n_e\gtrsim7.1\times10^8\,m_{\mu{\rm eV}}^2\,{\rm cm}^{-3}$ and energies between $\sim$ keV and MeV for emissions from the central engines of active galactic nuclei.
Such resonances can occur between the axion
mass term and the plasma frequency term as well as between the plasma frequency term
and the vacuum Cotton-Mouton shift. Assuming a limited region of production and conversion of order the gravitational radius we find that non-resonant effects can occur again in the keV and MeV energy range, provided that also the magnetic field is coherent over this distance. AGN jets and hot-spots should provide an even more interesting site with field strength of order $\mu$G--mG and jet sizes of order kpc. In these scenarios we find that a significant conversion is possible for MeV--TeV energies. Our findings should be testable in high energy $\gamma$-ray experiments. Additionally, though all our considerations encompassed normal axion parameters, our limits can be applied to axion-like particles like the PVLAS axion. In this case effects should even be observable down to the optical scale and thus be already testable. 

\begin{acknowledgments}
We thank G.~Raffelt for useful comments and discussions.
This work was partly supported by the Deutsche
Forschungsgemeinschaft under the grant TR-27 ``Neutrinos and
beyond'', the cluster of excellence for fundamental physics: Origin and Structure
of the Universe, and by the European Union under the ILIAS project, contract
No.~RII3-CT-2004-506222. KAH would like to thank the APC for hospitality.
\end{acknowledgments}


\begin{thebibliography}{99}

\bibitem{Battesti:2007um}
  R.~Battesti {\it et al.},
  ``Axion searches in the past, at present, and in the near future,''
  arXiv:0705.0615 [hep-ex].

\bibitem{Raffelt:1987im}
  G.~Raffelt and L.~Stodolsky,
  ``Mixing of the Photon with Low Mass Particles,''
  Phys.\ Rev.\  D {\bf 37}, 1237 (1988).

\bibitem{Mirizzi:2006zy}
  A.~Mirizzi, G.~G.~Raffelt and P.~D.~Serpico,
  ``Photon axion conversion in intergalactic magnetic fields and  cosmological
  consequences,''
  arXiv:astro-ph/0607415.

\bibitem{Mirizzi:2007hr}
  A.~Mirizzi, G.~G.~Raffelt and P.~D.~Serpico,
  ``Signatures of axion-like particles in the spectra of TeV gamma-ray
  sources,''
  arXiv:0704.3044 [astro-ph].

\bibitem{Adler:1971wn}
  S.~L.~Adler,
  ``Photon splitting and photon dispersion in a strong magnetic field,''
  Annals Phys.\  {\bf 67}, 599 (1971).

\bibitem{Erber:1966vv}
  T.~Erber,
  ``High-energy electromagnetic conversion processes in intense magnetic
  fields,''
  Rev.\ Mod.\ Phys.\  {\bf 38}, 626 (1966).

\bibitem{Zavattini:2005tm}
  E.~Zavattini {\it et al.}  [PVLAS Collaboration],
  ``Experimental observation of optical rotation generated in vacuum by a
  magnetic field,''
  Phys.\ Rev.\ Lett.\  {\bf 96}, 110406 (2006)
  [arXiv:hep-ex/0507107].
  
\bibitem{Zavattini:2007ee}
  E.~Zavattini {\it et al.}  [PVLAS Collaboration],
  ``New PVLAS results and limits on magnetically induced optical rotation and
  ellipticity in vacuum,''
  arXiv:0706.3419 [hep-ex].

\bibitem{m87-variability} F.~Aharonian et al.,
  ``Fast variability of tera-electron volt rays from the radio galaxy M87,''
  Science {\bf 314}, 1424 (2006).

\bibitem{Krennrich:2002as}
  F.~Krennrich {\it et al.},
  ``Discovery of Spectral Variability of Markarian 421 at TeV Energies,''
  Astrophys.\ J.\  {\bf 575}, L9 (2002)
  [arXiv:astro-ph/0207184].

\bibitem{Konopelko:2003zr}
  A.~K.~Konopelko, A.~Mastichiadis, J.~G.~Kirk, O.~C.~de Jager and F.~W.~Stecker,
  ``Modelling the TeV gamma-ray spectra of two low redshift AGNs: Mkn 501  and
  Mkn 421,''
  Astrophys.\ J.\  {\bf 597}, 851 (2003)
  [arXiv:astro-ph/0302049].

\bibitem{mrk501} J.~Kataoka et al.,
  ``High-Energy Emission from the TEV Blazar Markarian 501 during Multiwavelength
  Observations in 1996,''
  Astrophys.~J. {\bf 514}, 138 (1999).

\bibitem{Aharonian:2007nq}
  H.~E.~S.~S. collaboration, F.~Aharonian,
  ``Detection of VHE gamma-ray emission from the distant blazar 1ES 1101-232
  with H.E.S.S. and broadband characterisation,''
  arXiv:0705.2946 [astro-ph].

\bibitem{Cheung:2007wp}
  C.~C.~Cheung, D.~E.~Harris and L.~Stawarz,
  ``Superluminal Radio Features in the M87 Jet and the Site of Flaring TeV
  Gamma-ray Emission,''
  arXiv:0705.2448 [astro-ph].

\bibitem{Torres:2003zb}
  D.~F.~Torres,
  ``Gamma-ray sources at high latitudes,''
  arXiv:astro-ph/0308069.

\bibitem{Boettcher:2006pd}
  M.~Boettcher,
  ``Modeling the emission processes in blazars,''
  arXiv:astro-ph/0608713.

\bibitem{Hooper:2007bq}
  D.~Hooper and P.~D.~Serpico,
  ``Detecting Axion-Like Particles With Gamma Ray Telescopes,''
  arXiv:0706.3203 [hep-ph].

\bibitem{Aharonian:2005ti}
  F.~Aharonian and A.~Neronov,
  ``TeV gamma rays from the galactic center,''
  arXiv:astro-ph/0503354.

\bibitem{Aharonian:2004jr}
  F.~Aharonian and A.~Neronov,
  ``High energy gamma rays from the massive black hole in the galactic
  center,''
  Astrophys.\ J.\  {\bf 619}, 306 (2005)
  [arXiv:astro-ph/0408303].

\bibitem{Anchordoqui:2003vc}
  L.~A.~Anchordoqui, H.~Goldberg, F.~Halzen and T.~J.~Weiler,
  ``Galactic point sources of TeV antineutrinos,''
  Phys.\ Lett.\  B {\bf 593}, 42 (2004)
  [arXiv:astro-ph/0311002].

\bibitem{Dupays:2006dg}
  A.~Dupays and M.~Roncadelli,
  ``Discovering Light Pseudoscalar Bosons in Double-Pulsar Observations,''
  arXiv:astro-ph/0612176.

\bibitem{Dupays:2006hz}
  A.~Dupays and M.~Roncadelli,
  ``Light pseudoscalar bosons, PVLAS and the double pulsar J0737-3039,''
  arXiv:astro-ph/0612227.

\bibitem{vallee} J.~P.~Vall\'ee,
  ``Observations of the Magnetic Fields Inside and Outside the Milky Way,''
  Fundamentals of Cosmic Physics {\bf 19}, 1 (1997).

\bibitem{ohno} H.~Ohno, S.~Shibata,
  ``The random magnetic field in the Galaxy,''
  Mon.~Not.~R.~Astron.~Soc. {\bf 262}, 953 (1993).

\bibitem{han} J.~L.~Han,
  ``Magnetic fields in our Galaxy: How much do we know? III. Progress in the last decade,''
  Chin.~J.~Astron.~Astrophys. {\bf 6}, 211 (2006).

\bibitem{Mortsell:2003ya}
  E.~Mortsell and A.~Goobar,
  ``Constraining photon axion oscillations using quasar spectra,''
  JCAP {\bf 0304}, 003 (2003)
  [arXiv:astro-ph/0303081].

\bibitem{De Angelis:2007yu}
  A.~De Angelis, O.~Mansutti and M.~Roncadelli,
  ``Axion-Like Particles, Cosmic Magnetic Fields and Gamma-Ray Astrophysics,''
  arXiv:0707.2695 [astro-ph].

\bibitem{Camenzind:2004ae}
  M.~Camenzind,
  ``Relativistic Outflows form Active Galactic Nuclei,''
  arXiv:astro-ph/0411573.

\bibitem{magnetized-disk} G.~B.~Field and R.~D.~Rogers,
  ``Radiation from Magnetized Accretion Disks in Active Galactic Nuclei'',
  Astrophys.~J. {\bf 403}, 94 (1993).

\bibitem{kardashev} N.~S.~Kardashev,
  ``Cosmic Supercollider'',
  Mon.\ Not.\ R.\ Astron.\ Soc.\ {\bf 276}, 515 (1995).

\bibitem{Neronov:2007vy}
  A.~Neronov and F.~Aharonian,
  ``Production of TeV gamma-radiation in the vicinity of the supermassive
  black hole in the giant radiogalaxy M87,''
  arXiv:0704.3282 [astro-ph].

\bibitem{sikora} M.~Sikora, G.~Madejski, R.~Moderski, J.~Poutanen,
  ``Learning about Active Galactic Nucleus Jets from Spectral Properties of Blazars,''
  Astrophys.~J. {\bf 484}, 108 (1997).

\bibitem{sreekumar} P.~Sreekumar, AIP Conference Proceedings, {\bf 510}, 459 (2000).

\bibitem{ostrowski} L.~Stawarz, A.~Siemiginowska, M.~Ostrowski, M.~Sikora,
  ``On the Magnetic Field in the Kiloparsec-Scale Jet of Radio Galaxy M87,''
  Astrophys.~J. {\bf 626}, 120 (2005).

\bibitem{aharonian} F.~A.~Aharonian,
  ``Proton-synchrotron radiation of large-scale jets in active galactic nuclei,''
  Mon.\ Not.\ Roy.\ Astron.\ Soc.\  {\bf 332}, 215 (2002).

\bibitem{Stecker:2006vx}
  see, e.g., F.~W.~Stecker,
  ``Exploring the Edge of the Stellar Universe with Gamma-Ray Observations,''
  arXiv:astro-ph/0611455.

\bibitem{lobanov} A.~P.~Lobanov,
  ``Ultracompact jets in active galactic nuclei'',
  Astron.\ Astrophys.\ {\bf 330}, 79 (1998).

\end{thebibliography}
\end{document}